\documentclass{emulateapj}
\newcommand\gsim{\;\lower.6ex\hbox{$\sim$}\kern-7.75pt\raise.65ex\hbox{$>$}\;}
\newcommand\lsim{\;\lower.6ex\hbox{$\sim$}\kern-7.75pt\raise.65ex\hbox{$<$}\;}
\newcommand\mj{m$_{\mathrm{F110W}}$}
\newcommand\mh{m$_{\mathrm{F160W}}$}
\newcommand\mjh{m$_{\mathrm{F110W}}$-m$_{\mathrm{F160W}}$}
\newcommand{\Msun}{M$_{\odot}$}

\newcommand{\Myr}{M$_{\odot}$~yr$^{-1}$}
\newcommand{\Myk}{M$_{\odot}$~yr$^{-1}$~kpc$^{-2}$}

\shorttitle{The complex SFH of NGC 1569}
\shortauthors{Angeretti et al.}
\received{2004 August 10}
\accepted{2005 February 13}
\begin{document}
\title{The complex Star Formation History of NGC 1569 \thanks{Based 
on observations with the NASA/ESA Hubble Space Telescope, obtained 
at the Space Telescope Science Institute, which is 
operated by AURA Inc. for NASA under contract NAS 5-26555.}}

\author{L. Angeretti}
\affil{Universit\`a di Bologna -- Dipartimento di Astronomia, 
       via Ranzani, 1, I-40127 Bologna, Italy}
\email{luca.angeretti@bo.astro.it}

\author{M. Tosi}
\affil{INAF -- Osservatorio Astronomico di Bologna,
       via Ranzani 1, I-40127 Bologna, Italy}
\email{monica.tosi@bo.astro.it}

\author{L. Greggio} 
\affil{INAF -- Osservatorio Astronomico di Padova,
       vicolo dell'Osservatorio 5, I-35122, Padova, Italy}
\email{greggio@pd.astro.it}

\author{E. Sabbi}
\affil{Universit\`a di Bologna -- Dipartimento di Astronomia, 
       via Ranzani, 1, I-40127 Bologna, Italy}
\email{elena.sabbi@bo.astro.it}

\author{A. Aloisi\altaffilmark{1} \& Claus Leitherer}
\affil{Space Telescope Science Institute,
       3700 San Martin Drive, Baltimore, MD 21218, USA}
\email{aloisi@stsci.edu, leitherer@stsci.edu}

\altaffiltext{1}{On assignment from the Space Telescope Division of the European Space Agency}

%%%%%%%%%%%%%%%%%%%%%%%%%%%%%%%%%%%%%%%%%%%%%%%%%%%%%%%%%%%%%%%%%%%%%%%%%%%%%%%
%%                                                                           %%
%%                                                                           %%
%%                           A B S T R A C T                                 %%
%%                                                                           %%
%%                                                                           %%
%%%%%%%%%%%%%%%%%%%%%%%%%%%%%%%%%%%%%%%%%%%%%%%%%%%%%%%%%%%%%%%%%%%%%%%%%%%%%%%
\begin{abstract}
 We present new results on the star formation history of the dwarf irregular 
galaxy NGC 1569. The data were obtained with Hubble Space Telescope's  NICMOS/NIC2 
in the F110W (J) and F160W (H) near-infrared (NIR)  filters and interpreted with the 
synthetic color-magnitude diagram  method. 
 The galaxy experienced a complex star formation (SF) activity.
The best fit to the data is found assuming three episodes of activity in the last 
1 $-$ 2 Gyr. The most recent and strong episode constrained by these NIR data 
started $\sim 3.7 \times 10^7$ yr ago and ended $\sim 1.3 \times 10^7$ yr ago, 
although we cannot exclude  that up to three SF episodes occurred in this time interval.
The average star-formation rate (SFR) of the episode is
$\sim 3.2$ M$_{\odot}$ yr$^{-1}$ kpc$^{-2}$, in agreement with literature data.
A previous episode produced stars between $\sim 1.5\times 10^8$ yr and
$\sim 4 \times 10^7$ yr ago, with a mean SFR about 2/3 lower than the
mean SFR of the youngest episode. An older SF episode occurred about
$1 \times 10^9$ yr ago. All these SFRs are 2 $-$ 3 orders of magnitude higher
than those derived for late-type dwarfs of the Local Group.
In all cases an initial mass function similar to Salpeter's allows for a good
reproduction of the data, but we cannot exclude flatter mass functions.
These results  have been obtained adopting a distance 
of 2.2 Mpc and a reddening E(B-V)=0.56. A larger distance would require younger episodes 
and higher SFRs.
We have explored some possible scenarios using the astrated mass in the
best fit model, in order to constrain the past star formation history.
 We cannot rule out a low rate in the past SF but we can safely conclude
that the last 1 $-$ 2 Gyr have been peculiar. 
\end{abstract}

%%%%%%%%%%%%%%%%%%%%%%%%%%%%%%%%%%%%%%%%%%%%%%%%%%%%%%%%%%%%%%%%%%%%%%%%%%%%%%%
%%                                                                           %%
%%                                                                           %%
%%                          K E Y W O R D S                                  %%
%%                                                                           %%
%%                                                                           %%
%%%%%%%%%%%%%%%%%%%%%%%%%%%%%%%%%%%%%%%%%%%%%%%%%%%%%%%%%%%%%%%%%%%%%%%%%%%%%%%

%%%%%%%%%%%%%%%%%%%%%%%%%%%%%%%%%%%%%%%%%%%%%%%%%%%%%%%%%%%%%%%%%%%%%%%%%%%%%%%
%%                                                                           %%
%%                                                                           %%
%%                       I N T R O D U C T I O N                             %%
%%                                                                           %%
%%                                                                           %%
%%%%%%%%%%%%%%%%%%%%%%%%%%%%%%%%%%%%%%%%%%%%%%%%%%%%%%%%%%%%%%%%%%%%%%%%%%%%%%%
\section{Introduction}
In the last decade the field of galaxy evolution has gained substantial momentum,
owing to the wealth of high-quality multi-wavelength surveys from the main ground
and space-based telescopes and sophisticated theoretical models.
Nonetheless, the current knowledge of the major physical mechanisms for galaxy 
formation and evolution is still affected by serious uncertainties.
Traditionally, star formation (SF) in cosmology and galaxy formation studies is
naively described by simple universal laws, and the stellar initial mass function (IMF)
is assumed to be constant in time and space.
The only way to obtain detailed information on these crucial evolution parameters is to study 
the stellar populations of nearby galaxies and derive the SF history (SFH) over cosmological
timescales. One of the fundamental tools is the color-magnitude diagram (CMD) 
of the galaxies resolved into single stars. Since all the CMD features are related to the evolutionary 
state of the stellar populations, we can understand the SFH of a certain region within a galaxy by 
interpreting its CMD in terms of stellar evolution theory with the method of the synthetic 
CMDs. In theory, from the CMD we can infer how many bursts have occurred, their duration
and star formation rate (SFR), and the initial mass function \citep{tos91,gre98} for any 
star-formation law and metallicity. In practice, observational errors and 
theoretical uncertainties impose severe restrictions.
 The requirement of resolving a galaxy into stars limits 
the exploration of the SFH to nearby galaxies.
At present, even the largest telescopes only permit such studies for 
galaxies with distance $<$ 15 $-$ 20 Mpc, i.e., in the local universe. 
In this limited space volume we find many spirals and a plethora of both late and early dwarf
galaxies.

% Dwarfs
Dwarf irregular galaxies (dIrrs) and blue compact galaxies (BCDs) 
play a major role in the field of galaxy evolution \citep{grb97,mat98}.
With their high gas content and low metallicity, they  could be similar 
to primeval galaxies \citep{izo97} and the best possible
sites for determining the primordial $^4$He abundance \citep{izo04,oli04}.
It has been suggested that the excess of number counts of faint blue galaxies 
at redshift z $\sim$ 1 may be ascribed to dwarf galaxies undergoing their
first burst of SF\citep{ell97}. Dwarf galaxies are important in galaxy 
formation models since in hierarchical models they are assumed to be 
the building blocks from which other types of galaxies form through merging.

 Key questions on the evolution of late-type dwarfs are:
did all the dIrrs and BCDs experience SF in the past? 
If so, what are the properties of the SF episodes?
Several stellar population studies have suggested that dIrrs 
are characterized by a gasping (episodes of moderate activity separated by
short quiescent phases) SF
with no evidence of extended gaps \citep{tos91,sch01,dol03,ski03}. Their SFR is
not strong enough to explain the faint galaxy counts excess.

%% NGC 1569 parameters
% Distance
NGC 1569 is a dIrr near the Local Group. Its intrinsic
distance modulus has been estimated to be 26.71 $\pm$ 0.60 \citep{arp85,isr88,wal91}, corresponding
to a distance of 2.2 $\pm$ 0.6 Mpc. \cite{oco94} obtained a distance modulus of 27.0 $\pm$ 0.5
corresponding to 2.5 $\pm$ 0.5 Mpc, while \citet{kar94} and \citet{ric95} found 1.8 $\pm$ 0.4 Mpc 
and $\sim$ 1.7 Mpc, respectively. 
 Recently, \citet{mak03} derived a new distance to NGC 1569 based on the I magnitude of the tip 
 of the red giant branch (RGB). 
Their data are consistent with two possible distances: 1.95 $\pm$ 0.2 or 2.8 $\pm$ 0.2 Mpc. 
Hereafter we adopt a distance of 2.2 $\pm$ 0.6 Mpc \citep{isr88}, which encompasses the wide range of 
distances proposed in the literature.

% Reddening
 The low Galactic latitude (b $= 11.2^o$) of NGC 1569 implies significant Galactic 
reddening. \citet{bur84} estimated a value of $\mathrm{E(B-V)} = 0.51$ for the foreground extinction 
using the HI column density. \citet{isr88} found a total extinction of 
$\mathrm{E(B-V)} = 0.56 \pm 0.10$ from the UV color-color diagram.
Furthermore, many authors have suggested the existence of a strong, differential, internal reddening. 
\cite{gon97} found a total reddening of $\mathrm{E(B-V)} = 0.67 \pm 0.02$ in the direction 
of the super star clusters (SSCs),
where the component of the intrinsic reddening is  $\mathrm{E(B-V)} = 0.11$.
\cite{dev97} estimated the Galactic extinction to be $\mathrm{E(B-V)} = 0.50$ and the intrinsic mean 
extinction $\mathrm{E(B-V)} = 0.20$, using optical emission spectra of the ionized gas in  NGC 1569. 
\cite{kob97} found a variation of the total reddening in the galaxy from $\mathrm{E(B-V)} \simeq  0.51$
to $\mathrm{E(B-V)} \simeq  0.77$ with an average value $\mathrm{E(B-V)} \simeq 0.63$ (with R = 3.1). 
They found it reasonable to
assume a Galactic foreground of $\mathrm{E(B-V)} = 0.51$ and ascribe the difference to the internal reddening.
In \cite{kob97}, the extinction for the area discussed in the present paper is $\mathrm{E(B-V)} \sim 0.58$, 
except for the SSCs and HII regions.
\cite{ori01} derived a foreground extinction of $\mathrm{E(B-V)} = 0.55 \pm 0.05$ and
an intrinsic reddening of $\mathrm{E(B-V)} = 0.15 \pm 0.05$ from UV spectra. The simulations presented in this paper 
assume $\mathrm{E(B-V)} = 0.56 \pm 0.10$ \citep{isr88}.

% Metallicity
 The mean oxygen abundance of NGC 1569 found in the literature is $12+ \log(\mathrm{O/H}) \simeq  8.2 \pm 0.2$ 
dex. With an assumed [O/Fe]$=0.0$, the average metallicity is then 0.25 $Z_{\odot}$ corresponding to 
$\mathrm{Z} \sim 0.004$ \cite[see][and references therein]{gre98} if $\mathrm{Z}_{\odot} = 0.02$\footnote{In the last 
years, the solar abundances have undergone major revisions \citep{asp04}.
The current estimate of the solar metallicity is $\mathrm{Z}_{\odot,\mathrm{new}} = 0.013$}. 

% NGC 1569 in literature
 This rather exceptional dwarf galaxy is known to contain two SSCs, SSC-A and SSC-B,
with SSC-A having two components, A1 and A2 \citep{dem97,ori01}.
Many HII regions are observed in those zones where SF is currently active \citep{wal91}. 
Nevertheless, NGC 1569 is considered a ``post'' star-forming galaxy with a total dynamical mass of 
M$_{dyn}$~$\sim 3.3 \times 10^8$ \Msun \ \citep{isr88}, one--third of which is neutral hydrogen gas
\cite[M$_{HI}$ $\sim$ 1.3 $\times 10^8$ \Msun,][]{sti02}. 
This galaxy is a gas-rich system in a relatively early stage of chemical evolution.
Many filaments are found at different wavelengths, and there is a strong 
spatial correlation between the extended X-ray emission and the H$_{\alpha}$ 
filaments \citep{mar02}.
%% Radio
 The HI emission of NGC 1569 shows a dense, clumpy ridge distribution
surrounded by a more extended diffuse neutral hydrogen. There are 
discrete features such as arms and bridges \citep{sti02}.
%% IR
 In the IR, \citet{lin02} found both large and very small 
grains exposed to a strong radiation field. \cite{gal03} found different
dust properties in NGC 1569 (most of the grains are small) compared to other more 
metal rich galaxies. These results are consistent with the presence of shocks 
having dramatic effects on the dust. 
%% Opt
 In the optical, deep $H_{\alpha}$ images show filaments and arcs of warm 
ionized gas. These filaments have a length of several kiloparsecs with typical velocities around 
50 $-$ 100 km s$^{-1}$. The highest velocities, up to 200 km s$^{-1}$, have been 
attributed to expanding superbubbles with dynamical ages of 10 Myr \citep{wal91} 
generated by a strong SF episode.
%% X-ray
 At high energy, X-ray observations reveal that $\sim 60\% $ of the
soft X-ray emission in NGC 1569 is centered between the two SSCs. Its origin
is attributed to thermal emission from the hot gas of a galactic superwind
emanating from the disk \citep{hec95}.
Chandra observations find many high-mass X-ray binary systems and
large inhomogeneities in the metal abundances of the interstellar medium (ISM)
with ranges from $0.1 Z_{\odot}$ to $1 Z_{\odot}$ \citep{mar02}.
This is  evidence that the ISM has been affected by numerous supernova 
explosions  from a recent SF episode \citep{mar02}.
%% Hydrodinamical Simulation
Hydrodynamical simulations of \citet{der99} demonstrated that a galactic wind 
triggered only by SNII explosions (i.e. without stripping or other environmental
effects) would not be very effective in removing the ISM in NGC 1569. In fact, 
even if most of the metal-rich stellar ejecta 
can be efficiently expelled from the galaxy and dispersed in the 
intergalactic medium, cold and dense gas replenishes the central region of the galaxy 
a few hundred million years after the starburst.

 The SFH of NGC 1569 has most recently been studied by two 
 groups with Hubble Space Telescope (HST) data.
% Vallenari et al. 1996 outcomes
 In optical bands (V and I), \cite{val96} found a global SF episode from 100 Myr ago to 4 Myr ago and 
hints of an older episode of SF from 1.5 Gyr ago to 150 Myr ago. They ruled out the 
existence of long quiescent phases in the last 1.5 Gyr.
% Greggio et al 1998 outcomes
 Based on B and V data, \citet[hereafter G98]{gre98} found a global SF episode 
of $\gsim$ 0.1 Gyr duration, ending $\sim 5 - 10$ Myr ago. During the burst, the SFR was approximately 
constant, and, if quiescent periods occurred, they lasted less then $\sim$ 10 Myr.
The derived SFR is very high and equal to 3, 1 or 0.5 \Myr \ for a single slope initial mass 
function\footnote{Throughout this paper we parameterize the IMF as 
$\varphi(m) \propto m^{-\alpha}$. In this notation, Salpeter's slope is
$\alpha =$ 2.35} with an exponent $\alpha =$ 3, 2.6 or 2.35, respectively,  and a mass range 
from 0.1 to 120 \Msun.
% Aloisi et al. 2001 
From HST NICMOS images, \citet[hereafter AL01]{alo01} found that young stars 
(M $> 8$ \Msun, age $< 50$ Myr) are more clustered around the two SSCs, 
while the intermediate-age objects ($1.9$ \Msun \ $<$ M $<8$ \Msun, 50 Myr $<$ age $<$ 1 Gyr) have an almost
uniform distribution, and the old stars (M $< 1.9$ \Msun, age $> 1$ Gyr) are located at the 
outskirts of the starbursting region. 

% Goal?!?!
All these studies show that NGC 1569 experienced a strong SFR in recent epochs.
With a total mass of $\sim 3.3 \times 10^8$ \Msun, the system cannot have sustained 
such a strong activity over a Hubble time, since it would have consumed all the gas in 
less than 1 Gyr (G98). The question then is whether the SF in NGC 1569 has started only in
relatively recent epochs or has lasted for several gigayears but at much lower rates. 
Our goal is to obtain a better understanding of the past history of this galaxy. 
The best way to study old stellar populations, at least in principle, is to 
observe in the IR bands where the RGB phase is more 
evident, and where reddening effects are not as severe as in optical bands.

% Index
 In this work, we derive the SFH with the synthetic CMD method [\cite{tos91} 
 and G98] using the data of AL01 in the NIR filters. 
In Section 2 we  summarize the data and the observed CMD. 
In Section 3 we briefly describe the CMD method and the artificial star 
experiment in order to derive completeness and photometric errors. The results are presented 
in Section 4. They are discussed and compared with the literature in 
Section 5.

%%%%%%%%%%%%%%%%%%%%%%%%%%%%%%%%%%%%%%%%%%%%%%%%%%%%%%%%%%%%%%%%%%%%%%%%%%%%%%%
%%                                                                           %%
%%                                                                           %%
%%                 O B S E R V A T I O N S                                   %%
%%                                                                           %%
%%                                                                           %%
%%%%%%%%%%%%%%%%%%%%%%%%%%%%%%%%%%%%%%%%%%%%%%%%%%%%%%%%%%%%%%%%%%%%%%%%%%%%%%%
\begin{figure}[t!]
 \figurenum{1}
 \includegraphics[width=0.7\textwidth]{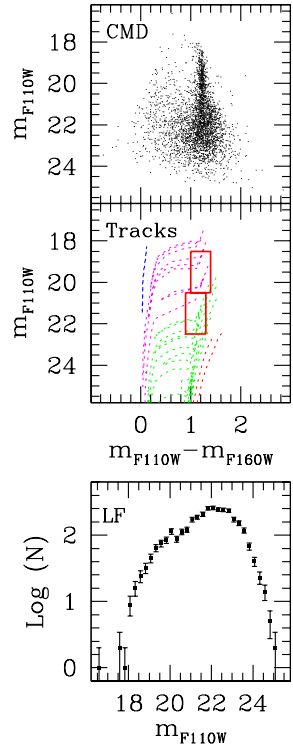}
 \label{CMDobs}
 \figcaption{Top panel: observed $m_{F110W}$ vs. $m_{F110W} - m_{F160W}$ color-magnitude diagram.
Middle panel: Padova tracks of metallicity $Z=0.004$ for 1, 1.5, 1.7, 2, 2.5, 3, 4, 5, 6, 9, 15, 20, 30 M$_{\odot}$
are shown as dashed lines. The boxes described in the text are plotted with solid lines: 
BOX1 (18.5 $-$ 20.5 and 1.0 $-$ 1.4) and 
BOX2 (20.5 $-$ 22.5 and 0.9 $-$ 1.3).
Bottom panel: the luminosity function with a step of $0.25$ mag.}
\end{figure}

\section{From the observations to the CMD}

% Observation and data reduction
We briefly summarize the main characteristics of the observations
and photometric reductions; for further details, see AL01.
 The observations of NGC 1569 were performed in 1998 with the NICMOS/NIC2 camera on board of 
HST in the F110W and F160W filters.
The images cover the crowded central region of the galaxy, where SSC-A, SSC-B, cluster \#30, and 
part of an extended HII region complex are present.
The NIC2 camera has a field of view of $19.2^{''} \times 19.2^{''}$, corresponding 
to $205 \times 205$  pc$^2$ at a distance of 2.2 Mpc.
The observations were taken with dithering.
The resulting ``drizzled'' images have a point-spread function (PSF) FWHM of 3.0 and 3.9 pixels 
in F110W and F160W, respectively, and a pixel size of $0.0375^{''}$, which
corresponds to $\sim 0.4$ pc.

The photometric reduction has been performed by AL01 using DAOPHOT via 
PSF fitting.
PSF-fitting photometry is not the ideal procedure for drizzled images,
because drizzle  creates a position-dependent PSF smearing and produces 
correlated noise between adjacent pixels \citep{fru02}. 
However PSF-fitting is the only viable 
technique in a crowded field like the central region of NGC 1569.
 AL01 used the deepest image (F110W) to locate the stars and used their 
coordinates to force photometry in the F160W image. This implies that all of the 
objects detected in the F160W image are also present in F110W but the opposite is not 
necessarily true. Then, they selected the objects by adopting the DAOPHOT parameters 
$\chi^2 < 1.5$ and sharpness $< 0.4$ and obtained the final catalog of 3177 stars.
The objects rejected by the selection criteria described above 
are typically the brightest objects in the SSCs, smaller star clusters, extended objects, and blends.
The star magnitudes were calibrated in the HST VEGAMAG system.

% The Color-Magnitude Diagram
 The CMD and luminosity function (LF) for NGC 1569 derived by AL01 are plotted in 
Figures~\ref{CMDobs}. The limiting magnitude is \mj \ $\simeq$ 24.5, and the color range is 
\mjh \ $= -0.8$ - 3. To easily identify the main evolutionary phases, 
we plotted the Padova tracks \citep{fag94} for a metallicity of Z = 0.004 on the same 
scale (Fig.~\ref{CMDobs}).
 The dominant feature is the ``red plume'' extending from \mj \ $\sim 18$ to $20.5$.
It has a small spread in color (0.2 mag) and a long range in magnitude. 
The region corresponds to the He burning phases of stars with masses 
approximately from 9 to 20 \Msun. The more massive stars in our catalog 
have masses of $15 - 20$ M$_{\odot}$, corresponding to ages of $10-15$ Myr. 
 A second feature is defined by the bulk of objects with \mj \ $\sim 21 - 24$ and \mjh \ $> 0.9$
where the tip of the RGB could be located.
 
Unfortunately, the RGB feature is not recognizable in the observed CMD.
 The tip of the RGB should be located at \mj \ $\simeq 22.4$ for an adopted
 distance modulus $\mathrm{DM} = 26.71$ and $\mathrm{E(B-V)} = 0.56.$
On the other hand, the DM uncertainty introduces  an uncertainty of 0.6 mag in the location
of the RGB tip (between 21.8 and 23.0 in \mj). In addition,
the RGB tip feature on the LF could be blurred in the CMD because 
the strong SF occurred at intermediate ages, as found by \citet{val96} 
and in this paper (see below).

%%%%%%%%%%%%%%%%%%%%%%%%%%%%%%%%%%%%%%%%%%%%%%%%%%%%%%%%%%%%%%%%%%%%%%%%%%%%%%%
%%                                                                           %%
%%                                                                           %%
%%                                   C M D                                   %%
%%                                                                           %%
%%                                                                           %%
%%%%%%%%%%%%%%%%%%%%%%%%%%%%%%%%%%%%%%%%%%%%%%%%%%%%%%%%%%%%%%%%%%%%%%%%%%%%%%%

\begin{figure}[t!]
 \figurenum{2}
 \includegraphics[scale=0.9, width=0.48\textwidth]{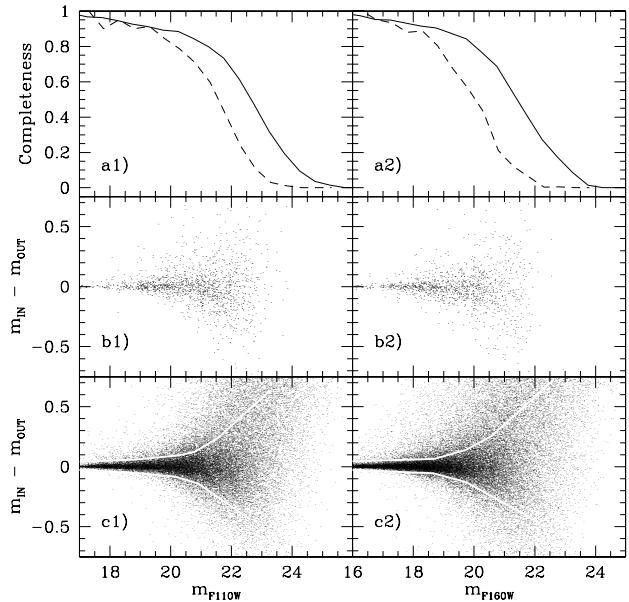}
 \label{Compl}
 \figcaption{The results of the artificial star experiments.
The left (right) panel shows the ASE results for the F110W (F160W) filter. 
The completeness functions are in the top panels. The solid lines refer to the completeness 
obtained in our study with Montegriffo's method described in \citet{tos01} and 
the dashed lines are from \citet{alo01} (hereafter AL01).
In the middle panels we show the photometric error distributions derived by AL01,
whilst the distributions obtained with Montegriffo's method are shown in the bottom panels. 
We indicate as thick lines the cuts in the error distributions adopted to perform 
the simulations presented in this paper.}
\end{figure}

\section{From the empirical to the synthetic CMDs}
 The derivation of the SFH from the CMD is critically dependent on the completeness
and analysis of the photometric errors. This is particularly true for the oldest 
populations in a crowded field such as the central region of NGC 1569, where
blending of unresolved stars may severely affect our data and 
create spurious features in the derived CMD.  It is therefore crucial to quantify 
these effects as well as possible by following the recipes of the data reduction,
and include them in the construction of the synthetic CMDs.

\subsection{Artificial star experiments}
 We accounted only for the completeness of the F160W frame in our simulations
because, as mentioned above, all the stars detected in F160W were previously
found in F110W. 
However, we have performed the artificial star experiment (ASE) in both images, since the 
photometric errors in both bands are required.

We initially adopted the ASE performed by AL01.
 They performed tests in which the fake stars have a PSF equal to that adopted in the data reduction. 
Then a fraction ($\sim$10\%) of the number of detected stars 
was added at random positions on the image in each run and about 10 runs 
were performed. The total number of artificial stars was then similar to that of the detected objects.
They derived an effective limiting magnitude of the data of \mh \ $\simeq 23$
and a 50\% completeness level at \mh \ $\simeq 20$.
 In the top panels of Figure~\ref{Compl} we plot their completeness and photometric error distributions.
The results obtained for \mh \ (\mj) are shown in the right (left) panel.
In the top panels we plot the completeness factors from AL01 as dashed lines, 
while in the 
middle panels we draw their estimated photometric error distributions. 

 When we applied the ASE results from AL01 to the simulations, 
 we found that they do not 
allow us to reproduce the faintest part of the CMD. For instance, if
we simulate an old SF episode (starting $5 \times 10^9$ yr ago 
and still active)
with the same number of stars of the observed CMD, we obtain a synthetic 
diagram that underproduces 
the faint (\mj \ $\gsim$ 23.5) tail of the LF by $50\%$ . 
We interpret this result as being due to an underestimate of the true completeness 
in the AL01 ASE, and we therefore performed new ASE computations. 
We believe that this is because AL01's approach does not add a
sufficient number of artificial stars  at faint magnitudes, where crowding,
blending and incompleteness are very severe. 
We therefore adopted a procedure developed by P. Montegriffo at the Bologna 
Observatory
\citep{tos01}, which uses $\sim 10^5$ artificial stars per band. We have put 
fake stars in the images 
according to a LF similar to the observed one. 
At brighter magnitudes this LF follows the observed one, but 
at low luminosities, a higher number of stars is required because of the high 
probability of losing objects because of crowding.
All the fake stars have a PSF equal to that adopted in the actual data reduction.
 To avoid artificial crowding, we divided the frames in grids of cells of 
known width, and we randomly positioned only one fake star per cell at each run.
We forced each star to have a distance from the cell edges 
large enough to guarantee that all its flux and background measuring regions fell
within the cell. In this way, we could control the minimum distance between adjacent stars. 
 At each run, the position of the grid is randomly changed and the artificial stars are
then searched in the image by using the same procedure adopted
for the real data and selected with the same criteria.

 An artificial star was considered lost when the difference between the input 
and output magnitude was 
larger than 0.75. Such a difference implies that the artificial star 
overlaps with another star of the same luminosity or brighter.
 In the top panels of Figure~\ref{Compl}, the completeness derived using Montegriffo's method
is shown as a solid line. The photometric error distributions are plotted 
in the bottom panels as dots.
The effective limiting magnitude of the data is \mh \ $\simeq 24.5$ %(\mj $\simeq 25$)
and the sample is 50\% complete at \mh \ $\simeq 21.5$. % (\mj $\simeq 22$).
 We found {\it a posteriori} that the completeness function obtained with Montegriffo's method 
allows us to reproduce the faint LF significantly better than with the previous ASE.

The input--output magnitude difference of the artificial stars provides an estimate of the
photometric errors, which is safer than the error indicators of the reduction package
($\sigma_{Daophot}$ in our case).  The outcomes are plotted in panels \ref{Compl}b and c
for the ASE approaches by AL01 and Montegriffo, respectively. 
 The synthetic CMDs obtained by adopting the photometric errors from  the artificial star 
tests, usually reproduce  the color and magnitude spread of the observational CMDs very well
[see e.g. \citet{gal02}, \citet{sme02}, \citet{tos02}, \citet{ann03}].
 In this case, however, by adopting the input--output magnitude difference of
Fig.~\ref{Compl}b as photometric error, we obtain an excessively large color spread in the sequences
of the synthetic CMD.  This effect is clearly evident along the red plume, where the observed CMD 
is tight in color. 
We therefore used the synthetic CMDs to heuristically infer the appropriate
error distribution in each filter. First, we cut the wings of the photometric 
error distributions at the $1\sigma$ level in each magnitude bin, and adopted these 
distributions to derive a test CMD.
Since in this test CMD the $1\sigma$ photometric error still has  a color 
distribution larger than observed,
we iterated the cutting procedure until we obtained photometric error 
distributions capable of reproducing the spread in the main features of the CMD.
The cuts performed on the photometric error distributions eventually adopted are
those shown within the solid lines in the bottom panels of Fig.~\ref{Compl}.

\subsection{Preliminary tests}
We have also checked whether different portions of the image have different 
completeness and photometric errors.  To this end, we have divided the NIC2 
images into three regions: the central one covering the most crowded part of the image 
including the SSCs, and the other two towards the edges in opposite directions.
 No relevant deviation has been found either in the completeness factors or in the
input--output magnitude difference of the three regions. 

Many authors have found evidence of spatially variable intrinsic extinction in 
NGC 1569; this extinction increases 
toward the central region of the galaxy, where SSCs and HII regions 
are located \citep{gon97,dev97,kob97,ori01}.
We have divided the image in subregions and compared the corresponding color functions 
in order to investigate whether strong differential reddening is present in our field.
No significant differences in the color distributions have been detected 
from the image center to the edges. We are confident that differential reddening
is not an issue in our $205 \times 205$ pc$^2$ sized region. 

 Since the CMDs of all the subregions show similar photometric errors, completenesses 
and color distributions, we have derived a global SFH over the whole $205 \times 205$ pc$^2$ image.

\subsection{The synthetic CMD method}
% The method
 The synthetic CMDs are constructed via Monte Carlo extractions 
of (mass, age) pairs, according to an assumed IMF, SF law,  and initial and final 
epoch of the SF activity.  Each synthetic star is placed in the theoretical 
$log(L/L_{\odot}), log(T_{eff})$ plane by suitable interpolations on the adopted 
evolutionary tracks \cite[in this case, the Padova sets by][]{fag94}. 
Luminosity and effective temperature are transformed into the desired 
photometric system by interpolation within appropriate tables for photometric 
conversions [the transformations to the VEGAMAG system by \citet{ori00}, 
in this case], and the resulting absolute magnitude 
is converted to a provisional apparent magnitude by applying reddening 
and distance modulus. An incompleteness test on the F160W magnitudes is then performed, based on 
the results from Montegriffo's ASE. 
A photometric error is assigned to all the retained stars by using the 
``modified distributions'' discussed in Section 3.1. 
The value of the SFR is obtained when the number of objects in the synthetic CMD 
(or in regions corresponding to specific evolutionary phases) is consistent with the 
observational data.
When the examined CMD region contains only a few objects, we perform several 
simulations in order to avoid problems due to small number statistics. 
In these cases, we adopted the average value of the resulting SFRs. 

%%%%%%%%%%%%%%%%%%%%%%%%%%%%%%%%%%%%%%%%%%%%%%%%%%%%%%%%%%%%%%%%%%%%%%%%%%%%%%%
%%                                                                           %%
%%                                                                           %%
%%                           S I M U L A T I O N S                           %%
%%                                                                           %%
%%                                                                           %%
%%%%%%%%%%%%%%%%%%%%%%%%%%%%%%%%%%%%%%%%%%%%%%%%%%%%%%%%%%%%%%%%%%%%%%%%%%%%%%%
\section{Simulations}
In order to derive the SFH of a galaxy, 
we explore a  parameter space dependent on the distance modulus, reddening, IMF, 
SF law, metallicity, and age of the episodes. We obtain a solution 
when we have a combination of parameters that generate a synthetic CMD similar to the 
observed one. The exploration is also limited by the intrinsic uncertainties of the 
theoretical models, like those introduced by the stellar evolution tracks and the
photometric conversions. 

First, we attempt to reproduce the observed CMD with a single episode and metallicity 
model over the whole range of ages covered by the empirical CMD.
If a single episode does not reproduce the observed CMD, we assume multiple SF
events. If a single metallicity does not allow us to simulate an
acceptable CMD, we try to change the metallicity for each episode.
Then we perform other simulations in order to check the effect on the
resulting SFH by assuming different IMFs, distance moduli and reddenings.

This approach gives us an indication of the uncertainty on the results of our 
simulations. It allows us to identify the range of the parameter values leading to
a satisfactory agreement with the data, and to reject those clearly inconsistent 
with the observational evidence. Following this line of thought, we will show the simulated 
episodes that fit the observed data but also some of the bad episodes, 
because they provide hints for a better understanding of the SFH. We
ran more than 1000 simulations to infer the SFH of the complicated case of NGC 1569.

 We  first present the results obtained by adopting $\mathrm{DM} = 26.71$, $\mathrm{E(B-V)} = 0.56$,
$Z = 0.004$ and a Salpeter IMF ($\alpha=$2.35) over the whole mass range from 0.1 to 120 \Msun.
In paragraph 4.3 we discuss the effect of assuming different values for the IMF slope,
distance and reddening parameters. The models all assume a constant SFR within the 
specific age ranges; we always report the SFR of each episode both in 
\Myk \ and in \Myr \ (the latter in brackets).

\begin{figure}
 \figurenum{3}
 \includegraphics[bb=18 270 385 718, width=0.48\textwidth]{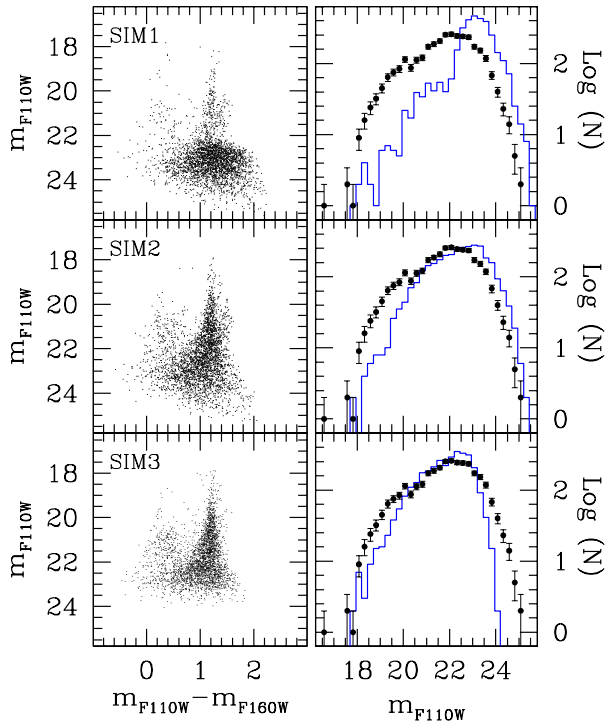}
 \label{Bad}
 \figcaption{Top panels SIM1: an old single-episode. The galaxy forms stars from 5 $\times$ 10$^9$ yr ago 
until today. The mean SFR is 0.28 \Myk.
Middle panels SIM2: a young single-episode. In this case NGC 1569 forms stars from 5 $\times$ 10$^8$ yr ago to
1 $\times$ $10^7$ yr ago with a mean SFR $\sim$ 1.1 \Myk.
Bottom panels SIM3: a single-episode that forms stars from 1.5 $\times$ $10^8$ yr ago to
1 $\times$ $10^7$ yr ago with a SFR of 1.7 \Myk.}
\end{figure}

\subsection{Single episode SF}
 We have run several cases with a single episode of SF in NGC 1569, 
all constrained to reproduce the total number of stars in the observed CMD. 
Here we describe only three representative models:
the first one (hereafter SIM1, plotted in Fig.~\ref{Bad})
assumes a long SF activity,  commencing 5 Gyr ago and still ongoing.
The second model (hereafter SIM2, plotted in Fig.~\ref{Bad}) assumes
a SF commencing only 500 Myr ago and terminating 10 Myr ago.
The third simulation (hereafter SIM3, plotted in Fig.~\ref{Bad}) generates stars
from 150 to 10 Myr ago.

% Bad simulation 1 
SIM1 has a SFR of 0.28 \Myk \ (0.01 \Myr).  The simulated objects are 
concentrated in the lower, fainter part of the CMD (\mj \ $\gsim$ 22.5),
where the low-mass stars in the RGB phase are located. The observed CMD does 
not show a similar concentration of objects in the faintest part. Conversely,
at brighter magnitudes (\mj \ $<$ 22.5), the synthetic red plume shows a star 
deficiency with respect to the observed LF. These inconsistencies show that
the bulk of the assumed SF activity is located at excessively early epochs
and provides too many old stars.
 At the same time the synthetic CMD is populated by blue massive stars (with M $> 20$ \Msun). 
 A dozen of these synthetic stars are in the main-sequence phase.  %%(\mj $>20.5$ and \mjh $<0.9$).
However, we cannot clearly distinguish the main-sequence from the other evolved phases 
in the observed CMD because of the large photometric errors at faint magnitudes.
 We have also performed several simulations to overcome the small number statistics of 
the stars located at bright luminosity (\mj \ $< 18.5$). 
These simulations suggest that too many red supergiant stars are produced 
without observational counterparts.
We can conclude that, in addition to the excess of old stars in the synthetic CMD, we also 
have an excess of young stars. This is the reason why in SIM2 we have turned on the SF more recently and 
turned it off earlier (10 Myr ago).

% Bad simulation 2
SIM2 has a SFR $\sim$ 1.1 \Myk \ ($\sim$ 0.04 \Myr) and fails to account for the 
brighter stars of the red plume. Again, at faint magnitudes, SIM2 shows an 
excess of stars. In both simulations (SIM1 and SIM2), we have a feature in the 
region $19.5 <$ \mj \ $< 21.5$ and $1.45 <$ \mjh \ $< 1.7$, which corresponds to stars 
with masses $4 - 6$ \Msun \  in the AGB phase and does not appear in the observed CMD. 
We then tried to simulate the CMD with a single episode starting $1.5 \times 10^8$ yr ago 
and terminating 10 Myr ago in order to avoid these stars.
%Bad Simulation 3
SIM3 has a SFR of $1.7$ \Myk \ (0.07 \Myr). This simulation avoids the feature obtained
in SIM1 and SIM2 but shows a shortage of stars at low luminosities (\mj \ $> 23$) and
at the top of the red plume (\mj \ $<20$), and an excess of objects in the 
middle ($20<$ \mj \ $<23$) of the luminosity function. 

At the end of these preliminary simulations, no solution has been found that simultaneously fits 
both the bright and the faint parts of the CMD. 
Hence, a single episode cannot reproduce the observed CMD: to improve the fit to the data, 
we need to perform multiple-episode cases.
 
\begin{figure}
 \figurenum{4}
 \includegraphics[width=0.48\textwidth]{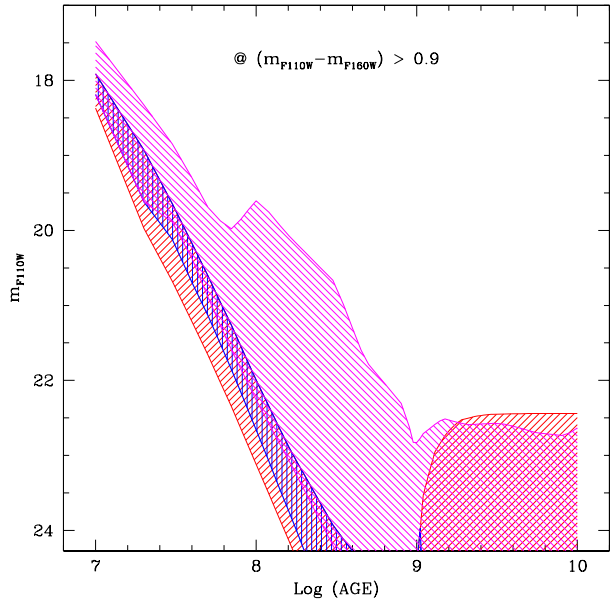}
 \label{Box}
 \figcaption{Magnitude range covered by red post-main-sequence stars of a simple stellar
population with an age between 10 Myr and 10 Gyr selected in color \mjh \ $>$ 0.9. 
Vertical shading refers to He-burning stars; slanted shading to RGB 
(angle = +45) and E-AGB (angle = -45) stars. 
This plot clearly shows that the temporal resolution is worse at older
ages.} 
\end{figure}
\subsection{Multiple episodes SF}
 We use the age-box procedure to investigate multiple episodes.
The basic idea is that we can perform our analysis by dividing the CMD in regions (boxes) populated 
by stars with selected properties, such as age or evolutionary phase \citep{gre02}.
A similar procedure is adopted by \cite{gal96}.
The star counts in each box are directly linked to the mass that went into stars
in the SF episodes that occurred at the corresponding epochs, i.e. the average 
SFR over the specific time intervals \citep{gre02}. 

In order to better understand the age-box procedure, we introduce Figure~\ref{Box} where 
we plot the apparent magnitude versus age of red (\mjh \ $> 0.9$) post main sequence stars. 
 Vertical shading indicates to He-burning stars, and slanted shading indicates  RGB (angle $= +45^o$) and 
E-AGB (angle $= -45$) stars. Because of the evolutionary lifetimes, the red stars at magnitudes brighter 
than 22.4 (where the tip of the RGB is located) are mostly core-He burning objects. The straight 
relation between their magnitude and age (Fig.~\ref{Box}, thick shaded strip) 
translates into a good temporal resolution of the SFH at ages younger than about 0.1 Gyr.
This allows us to obtain precise results in the boxes corresponding to the top of the red plume.
At magnitudes fainter than 22.4, each LF bin collects (in principle) stars with
ages up to the Hubble time. The stellar counts in this region allow
us to derive only integrated information.

% Single box
In practice, we first define the magnitude ranges of the box by identifying
the features in the LF\footnote{The comparison between model and data is performed 
selecting the red LF. However, in this paper we report only on the total luminosity function, 
since our CMD is mostly composed by a red population and the red LF does not significantly differ  
from the total one.}  that suggest a possible discontinuity in the star formation.
The color ranges are chosen in such a way to account for the spread because of the photometric errors. 
These ranges constrain the box that defines our partial model,
which is constrained to reproduce the observed stellar counts in the box with a 
SF activity within a specific age range. The limits for the age range are read from Fig. \ref{Box}
in correspondence to the magnitude limits of the box. Besides the objects in the box,
the simulation also populates the diagram in regions different from the box described above; 
we take these ``additional'' objects into account when we perform the simulations in 
the other parts of the CMD. 
We have run hundreds of cases to infer the best combination of parameters.
However, for the sake of clarity, here we simply describe the line of reasoning
and show the most relevant results from all these tests.
This method has been applied and extensively discussed by \cite{sch00,cro02,sch02,ann03}.

\subsubsection{One box simulations}
% Box 1
 Following the above reasoning, we define the box (hereafter BOX1) 
in the magnitude range $18.5 <$ \mj \ $< 20.5$ (at \mj \ $= 20.5$ a dip in the LF is noticeable) 
and colors $1.0 <$ \mjh \ $< 1.4$ (to take into account the photometric errors).
 Figure~\ref{Box} shows that in BOX1 we are sampling core-He burning
stars in the age range $(1 - 4) \times 10^7$ yr; these objects are the youngest, most massive 
stars of the red plume. This box also includes a population of lower-mass stars in the AGB phase; 
because of their short lifetimes in this magnitude range, these stars provide a minor 
contribution to the overall population.

\begin{figure}
 \figurenum{5}
% \epsscale{0.85}
% \plotone{f5.ps}
 \includegraphics[bb=18 143 385 718, width=0.48\textwidth]{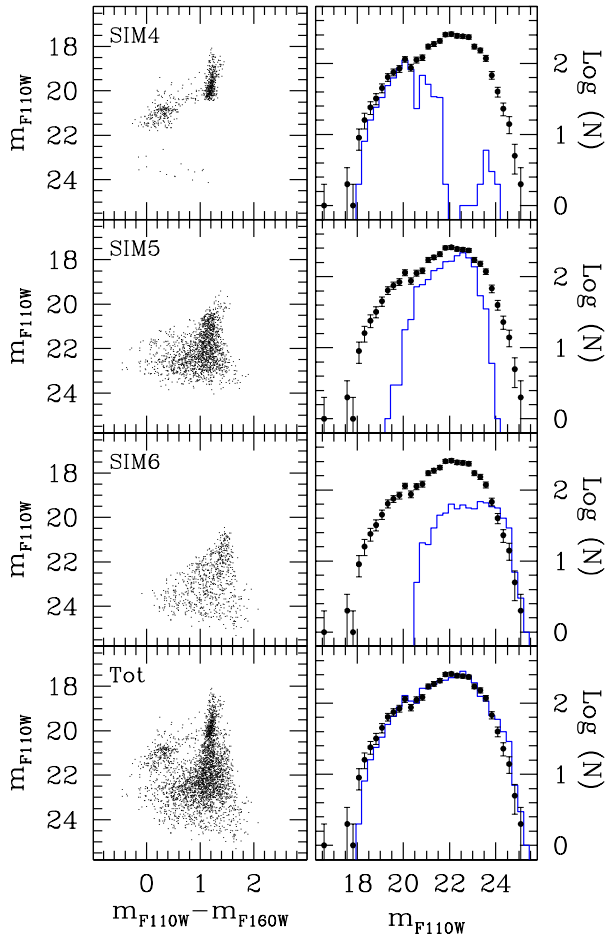}
 \label{Best}
 \figcaption{The best synthetic CMDs and luminosity functions.
In the top panel the youngest episode (SIM4) that reproduces the objects in BOX1 is plotted. 
The galaxy experienced strong SF beginning  3.7 $\times$ 10$^7$ yr ago and terminating 
1.3 $\times$ 10$^7$ yr ago with a high rate of $\sim$ 3.2 \Myk.
In the second panel from the top we plotted the intermediate episode (SIM5) constrained by BOX2. 
This episode occurred in the interval $1.5 \times 10^8 - 4 \times 10^7$ yr ago with a SFR $\sim$ 1.2 \Myk. 
The oldest episode that reproduces the remnant objects (SIM6),
is reported in the third panel from the top. Age interval: 6 $\times$ 10$^8$  $-$  3 $\times$ 10$^8$ yr ago.
SFR = 1.5 \Myk. The final composed synthetic CMD is drawn in the bottom panel.}
\end{figure}

After many simulations, we find the best parameter combination (SIM4) that 
reproduces the number counts in BOX1 (Fig.~\ref{Best}). 
SIM4 is an episode commencing $3.7\times 10^7$ yr ago and finishing 
$1.3 \times 10^7$ yr ago with a mean SFR of $3.1 - 3.3$ \Myk \ ($0.12 - 0.14$ \Myr).
We performed other simulations by slightly changing the boundary ages of 
the episodes for an estimate of the uncertainty affecting the determination
of the age intervals.

Other possible solutions have a starting epoch in the range $(3.5 - 3.8) \times 10^7$ yr ago,
and an ending epoch in the range $(1.3 - 1.4) \times 10^7$ yr ago. In all the
explored cases, 
the simulations for this young episode share a problem that is also visible in SIM4:
an overproduction of stars with \mjh \ $<0.5$ (hereafter {\it blue 
bulk}). This  structure corresponds to the blue edge of the blue 
loop phase of $8 - 12$ \Msun. We cannot remove this feature without 
underproducing the objects in the red plume.
The number of objects generated in this recent SF episode is 20\% of the total. 

We have then tried to simulate the remaining objects with a single episode, but we have not  succeeded
in reproducing the LF. If we construct single episodes starting at 
epochs earlier than $3 \times 10^8$ yr ago
and terminating $4 \times 10^7$ yr ago, we obtain too many faint objects (\mj \ $>22.5$) 
and too few stars in the range \mj \ $20.5 - 22.5$ in the synthetic diagram. 
On the other hand, an episode that begins to form stars $\lsim 3\times 10^8$ yr ago, 
does not produce a sufficient number of objects at faint magnitudes.

The lack of success in reproducing the LF with only two episodes led us to consider other scenarios.

\subsubsection{Two box simulations}
% Box 2
 The second box (hereafter BOX2) is defined in the magnitude range  $20.5<$ \mj \ $< 22.5$
and color $0.9<$ \mjh \ $<1.3$ and targets the intermediate ages.
The bright limit is fixed by BOX1, while the faint limit is chosen to be slightly
brighter than the RGB tip in order to exclude possible blended RGB stars from
the counts. From Fig.~\ref{Box}, one can see that stars falling in BOX2 have ages in the 
range $1.2 \times 10^8$ yr to $4 \times 10^7$ yr, if they are core-He burners. 
The color limits of BOX2 were indeed chosen with the aim of isolating core-He burning stars, 
while taking into account the color spread due to photometric errors.

The best simulation for BOX2 (hereafter SIM5) accounts for $\sim 50\%$ of the objects
and is plotted in the second panels from the top in Fig.~\ref{Best}. 
SIM5 has a SFR $=$ 1.2 \Myk \ (0.04 \Myr). 
This episode started forming stars $1.5 \times 10^8$ yr ago and terminated $4.0 \times 10^7$ yr ago.
 The starting epoch is older than the $1.2 \times 10^8$ yr mentioned
above. When computing the simulations, we vary the starting and
ending epoch around the values derived from Fig.~\ref{Box} in order to fit the observed 
LF with the minimum number of individual SF episodes. It turns out that
starting at $1.5 \times 10^8$ yr allows us to fit the stellar counts in the range
22.5 $-$ 23.5 without adopting a new box. Notice that at such magnitudes the
completeness factors fall below $50\%$.  

If we assume that a quiescent phase (or one with low SFR) occurred $(3.7 - 4) \times 10^7$ yr ago, 
we can also reproduce the shape of the LF in the magnitude range \mj \ $20 - 20.5$.
This kind of short inter-burst phases has also been found by G98.

% Others episodes
\subsubsection{The remaining objects}
The simulations computed to reproduce the stellar counts in BOX1 and BOX2
account for about 70\% of the total number of observed objects. 
 The remaining 30\% must be ascribed to SF that occurred at epochs 
 older than $1.5 \times 10^8$ yr. 
From Fig.~\ref{Box}, we see that the CMD in the magnitude range $22.5 - 24.5$ can 
be populated with a mixture of relatively young He-burning stars 
[with ages of $(1.5 - 3) \times 10^8$ yr], early-AGB stars with intermediate ages
(up to 1 $\times 10^9$ yr), and RGB stars with ages up to the Hubble time.
Therefore, we tested various starting epochs for the oldest episode of SF
sampled by our data.

First, we attempted to account for the remaining stars by considering He-burning
objects with ages between $3\times 10^8$ yr and $1.5 \times 10^8$ yr. The simulation fits 
 the LF fainter than \mj \ $ < 22.5$ reasonably well, but yields 
too many stars in the region $19.5 <$ \mj \ $< 21$ at colors $1.45 <$ \mjh \ $ < 1.7$.
The overproduction amounts to a factor of $4 - 5$. These are bright 
E-AGB stars with masses between 4 and 6 \Msun. As discussed in Section 4.1,
these synthetic objects were also produced in SIM2, which assumed active
SF at intermediate ages. 
If we use the number counts in the box $19.5 <$ \mj \ $<21$; $1.45 <$ \mjh \ $<1.7$ 
to constrain the SF activity at ages between $1.5\times 10^8$ and $3\times 10^8$ yr ago,
we obtain a SFR level of 0.14 \Myk \ (0.006 \Myr), which is 1 order of magnitude lower 
than the rate of NGC 1569 in the young and intermediate episodes.
We then tested SF episodes setting in at various epochs, but all ending $3 \times 10^8$ yr
ago, to avoid the overproduction of these bright AGB stars.
We found that if the old SF episode started a few Gyr ago, the 
upper RGB stands out very clearly on the synthetic CMD, with a prominent
horizontal feature corresponding to the tip. This is not seen in the data.
For a starting epoch of $2 \times 10^9$ yr ago, the signature of the RGB tip is no longer
apparent, and we obtain consistency with both the CMD and the LF.

We obtain acceptable models for any starting epoch in the range 2 -- 0.6 Gyr,
with corresponding variations of the SFR. For example, a simulation started
$1.5 \times 10^9$ yr ago yields a SFR of 0.4 \Myk \ (0.009 \Myr);
a simulation started $6 \times 10^8$ yr ago yields SFR $=$ 1.5 \Myk \ (0.06 \Myr). 
Our best fit is actually the latter, since the former case
tends to underproduce the faint AGB stars, corresponding to 3 \Msun \ objects ($20.5 >$\mj \ $> 21.5$) .
The best simulation for the remaining objects is shown in Fig.~\ref{Best} as SIM6, while  
the best composite CMD is plotted in Fig.~\ref{Best} with the label {\it Tot}.

\begin{figure}
 \figurenum{6}
 \includegraphics[width=0.48\textwidth]{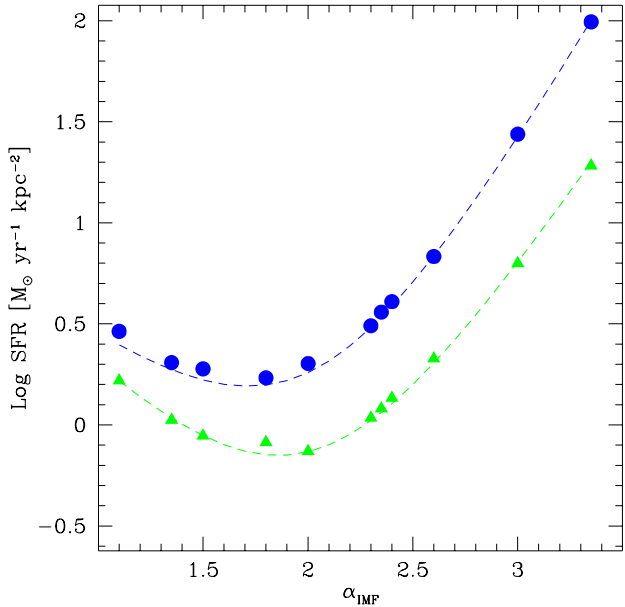}
 \label{IMFSFR}
 \figcaption{The trend of the SFR as a function of the IMF slope ($\alpha$).
Filled circles refer to the youngest episode (BOX1, 8 \Msun \ $<$ M $<$ 15 \Msun) 
and filled triangles to the intermediate one (BOX2, 5 \Msun \ $<$ M $<$ 8 \Msun).
Dashed lines show the theoretical relation between the IMF slope and the SFR
of the corresponding box. The differences are due to stochastic fluctuations.}
\end{figure}

\subsection{Searching for other parameter combinations}
%% Effects of IMF
We have performed many other simulations in order to search for alternative solutions; 
in particular, we considered cases with different IMFs, with slope $\alpha$ in the interval $1.35 - 3.35$. 
Fig.~\ref{IMFSFR} plots the SFR as a function of the IMF exponent, for the
youngest (circles) and intermediate (triangles) episodes. 
Fig.~\ref{IMFSFR} also shows that the SFR has a minimum at $\alpha_{min} \simeq 1.5 - 2$.
 This behavior is due to the non-linear relation between the slope of the IMF 
and the number of stars per unit mass of the parent stellar population. 
By adopting a steep IMF (e.g. $\alpha \simeq 3$), the low-mass stars are relevant in number 
and weight with respect to the
total population, while the contribution of high-mass stars to the total mass is negligible.
 For a flat IMF (e.g. $\alpha \simeq 1.3$), low-mass stars are still numerous, 
but their weight decreases. 
On the other hand, the number of massive objects become slightly higher but 
their contribution to the total mass increases more than the total number. 

 We can broadly reproduce the shape of the LF in BOX1 with an IMF flatter than 
 Salpeter's 
(down to $\alpha = 1.35$). For simplicity, in this paper we have adopted a single-slope IMF, extending 
Salpeter's IMF to the mass range $0.1 - 120$ \Msun. 
\citet{gou97} suggest a slope flatter than Salpeter's ($\alpha = -0.56$) for M $< 0.6$ \Msun.
\cite{cha03} has also found that the IMF in the Galaxy depends weakly on the environment 
and is well described by a power-law for M $\gsim 1$ \Msun \ and a log-normal form below
this mass, except possibly for early SF conditions.
For the Galactic disk, \citet{kro03} found an IMF with a slope $\alpha = 1.3$ in the range 
$0.08 - 0.5$ \Msun \ and a slope that is consistent with Salpeter's above $0.5$ \Msun.  
If we extrapolate the SFR at $M < 0.5$ \Msun \ by adopting Kroupa's rather 
than Salpeter's IMF, the values become $30\%$ smaller than with the default Salpeter IMF.

% 3Myr Gap 
We have performed quantitative comparison ($\chi^2$) between our best model 
and alternative solutions without the gap in the age range 37$-$40 Myr, and
found that the model with the gap is $\sim 2000$ times more probable than the 
fit without the gap.

% Complex recent SF
Another possible SFH reproducing the brightest portion 
of the red plume consists of many short episodes in the age interval $(1 - 4) \times 10^7$ yr ago. 
This possibility is also indicated by the ``granularity'' of the LF. 
In this case we obtain the best fit with three episodes
separated by short periods (few million years) of quiescence or very low SFR.  
SIM4a started $2.1\times 10^7$ yr ago and terminated $1.3 \times 10^7$ yr 
ago with a SFR of $\sim 4.2$ \Myk \ ($\sim 0.18$ \Myr);
SIM4b started $3.0 \times 10^7$ yr ago and terminated $2.2 \times 10^7$ yr 
ago with a SFR of $\sim 4.5$ \Myk \ ($\sim 0.19$ \Myr);
SIM4c started $3.7 \times 10^7$ yr ago and terminated $3.3 \times 10^7$ yr
ago with a SFR of $\sim 3.2$ \Myk \ ($\sim 0.13$ \Myr).
This SFH reproduces remarkably well the brightest portion of the total LF.
 These three episodes have SFRs equal or higher than the single episode (see SIM4)
but their total astrated mass is similar to that.
We prefer not to pursue this fine tuning (although we cannot 
exclude a more complex scenario in the recent star formation) because the shape
of the red plume and its LF may be affected by residual star-like systems
mistaken as individual stars in spite of all our checks (see AL01) .

%% Effects of Distance and reddening
  There are no clear evolutionary features in the observed CMD (Fig.~\ref{CMDobs}) 
that allow a precise estimate of the distance modulus of the galaxy.
Therefore, we have performed several simulations in order to examine the effects of changing 
the distance on the derived SFH in NGC 1569. To this end,  we have adopted 
the lowest and the highest values of the DM found in the literature ($\mathrm{DM} = 26.1 - 27.3$).
 In general,  we obtain a different fit to the data by changing the DM.
For sake of clarity, we only discuss the case of the larger distance, since 
this could be important in order to understand whether the RGB tip can be hidden.
 In this case, we derive a younger episode with higher SFR with respect to our best model. 
This is because a given apparent magnitude samples intrinsically brighter, and hence more massive,
stars when DM is larger.   As a consequence, a larger number of massive and young stars 
needs to be reproduced by the simulations.
 Also, when we modify the distance of the galaxy, we also change the position of the evolutionary 
features of the synthetic CMD. This effect is important because it could substantially modify the SFH. 
 By adopting $\mathrm{DM} = 27.3$, the best fit is obtained with a young episode that has formed 
objects in the age interval $2.7 \times 10^7 - 8 \times 10^6$ yr ago, and the intermediate SF 
occurred  $1 \times 10^8 - 3.2 \times 10^7$ yr ago.
The corresponding SFRs are about 2 times higher than those found for $\mathrm{DM} = 26.71$ in SIM4 
and SIM5. A SF with a rate of 0.8 \Myk \ (0.04 \Myr) that occurred in the age interval 
$1.3 \times 10^9 - 2 \times 10^8$ yr ago is consistent with the data.
We can exclude a strong SF occurring at ages older than 1 $-$ 2 Gyr
because the RGB tip would be clearly present in the CMD despite the higher distance.
We want to stress again that, for what concerns the older SF, the results must be qualified 
by considering the uncertainties at faint magnitudes because of the incompleteness and 
photometric errors.

The reddening uncertainty suggested by the spread of the literature values shifts the synthetic CMD 
by 0.05 in color (\mjh) and 0.1 in magnitude (\mj); its effect on the age of the episodes is 
smaller than those due to the uncertainties on the distance modulus and tracks. 

%% Effects of the metallicity
Finally, with the synthetic CMD method, we can provide hints on the metallicity of the SF episode.
Our capability to select the metallicity of an episode depends on the photometric errors 
relative to its main evolutionary phase. In principle, we are able to reject all the metallicities 
that create unobserved features in the synthetic CMD. 
The colors of the red plume are best reproduced with Z=0.004; the Z=0.0004
tracks yield red supergiants that are too blue compared to the observations.
The Z=0.008 tracks instead produce very red objects.
At fainter magnitudes, where the photometric errors are larger, the presence
of a low Z population cannot be ruled out, while the Z=0.008 tracks
produce stars that are too red with respect to the data.

\subsection{Observational and theoretical uncertainties}
The low galactic latitude of NGC 1569 ($b \sim 11^o.2 $) led us to check 
the number of  Milky Way stars falling into our field of view. 
AL01 ran the Galaxy model described by \cite{cas90} and found that
the foreground contamination in our (small) analyzed region is negligible. 
Major problems in our analysis are instead crowding, the shallow limiting
magnitude and the high reddening. Crowding is severe in the central region 
because the high recent SF activity has generated many bright, young objects 
that hide fainter, older ones.
In addition, the shallow limiting magnitude does not allow us to reliably 
detect features that could help us to better determine the characteristics 
of the older episode. From this point of view, despite the high reddening,
deeper, higher resolution HST images in optical bands could have possibly 
provided a better data set to study the earlier epochs. This has been clearly demonstrated 
by \cite{ann03} in NGC 1705 (a BCD twice as distant as NGC 1569). In that case, WFPC2
photometry has allowed to reach $1- 1.5$ mag below the RGB tip in regions
where NIC2 barely allows the identification of the tip. 

%% Clusters
An additional uncertainty on the analysis of the central region of NGC 1569 
comes from the presence of many  small clusters \citep{hun00,and04} and 
star-like systems. Most of them appear as {\it bona fide } stellar objects in our photometry, 
with shape, sharpness, and $\chi^2$ parameters typical of single stars. 
These unresolved systems could be located in the CMD at the top of the red 
plume. By keeping or removing all possible cluster candidates from our
empirical CMD, we estimate that the star-like systems could change the 
SFR by 15 $-$ 20\% and shift the end epoch of the youngest episode 
to an age a few Myr older.

\begin{figure}
 \figurenum{7}
 \includegraphics[bb=18 520 385 718, width=0.48\textwidth]{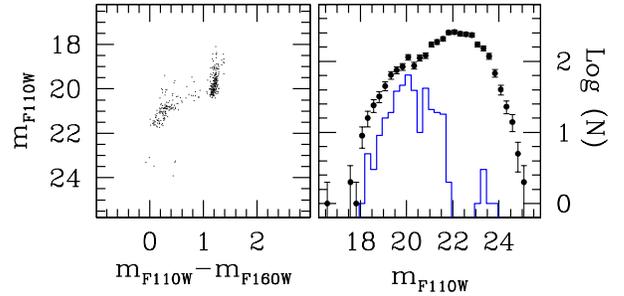}
 \label{BB}
 \figcaption{Simulation attempting to reproduce the number of objects 
in the box of the {\it blue bulk}. The number of objects 
at the top of the red plume is underestimated: 
this is an indication that the SFR found in SIM4 is an upper limit.}
\end{figure}

% Problems in Theoretical Tracks??
%% Blue  bulk
As discussed above and shown in the top panels of Fig.~\ref{Best}, what 
we called {\it blue bulk} in
Sect.~4.2 could affect the derivation of the SFR of the youngest episode.
This spurious feature could be due either to uncertainties in the bolometric 
corrections in the infrared filters or to uncertainties in the stellar 
tracks at the blue edge of the blue loops for the masses 8 $-$ 12 M$_{\odot}$.
In order to evaluate the influence of the {\it blue bulk}, we re-simulated the top of 
the red plume by locating a  box in the region of the {\it blue bulk} 
(20.5 $<$ \mj \ $<$ 21.5 and 0.9 $<$ \mjh \ $<$ 1.3).
By using the same age interval of the youngest episode, the simulation generates 
a partial model, plotted in Fig.~\ref{BB}, which provides a SFR of 
1.6 \Myk \ (0.06 \Myr). This simulation is able to generate only 50\% of the 
total objects in the red plume of the observed CMD and is clearly inconsistent
with the data.  We also performed many simulations by using Padova tracks of different 
metallicity, but we found similar problems.
In other words, either we reproduce the number of stars in the
{\it blue bulk} region, but severely underproduce the red plume objects, or we
reproduce the red plume and overestimate the {\it blue bulk}. We consider the latter
option more viable, since the timescales of stellar evolution models at the
blue edge of the blue loops are more uncertain than those of the red plume phases. 
Notice that  \citet{gre98} have derived the SFH in the B and V bands adopting both 
the Geneva and Padova
tracks in order to evaluate the uncertainty related to different stellar 
evolutionary
models. They found an excess of synthetic stars at the blue edge of the blue 
loop with the Geneva tracks 
but not with the Padova tracks. This suggests that an additional uncertainty 
comes into play in the {\it blue bulk}; 
for instance, the completeness factor could be overestimated because 
its color dependence has not been considered in the 
simulations.

%%%%%%%%%%%%%%%%%%%%%%%%%%%%%%%%%%%%%%%%%%%%%%%%%%%%%%%%%%%%%%%%%%%%%%%%%%%%%%%
%%                                                                           %%
%%                                                                           %%
%%                   D I S C U S S I O N  &  C O N C L U S I O N             %%
%%                                                                           %%
%%                                                                           %%
%%%%%%%%%%%%%%%%%%%%%%%%%%%%%%%%%%%%%%%%%%%%%%%%%%%%%%%%%%%%%%%%%%%%%%%%%%%%%%%

\begin{figure}
 \figurenum{8}
 \includegraphics[width=0.48\textwidth]{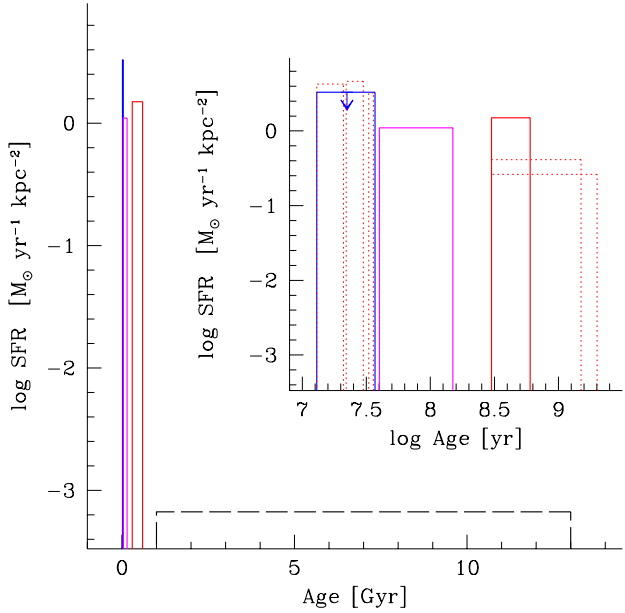}
 \label{sfh}
 \figcaption{The derived SFH of NGC 1569. The SFH over the Hubble time is reported in the main 
plot linearly in age. 
The solid lines refer to the episodes of Figure~\ref{Best} and the dashed line
to the second scenario described in the conclusions.
The recent star formation history resulting from the 
simulations is shown in the small logarithmic plot.  We give only an upper limit for the youngest episode. 
These uncertainties in the SFR are due to the {\it blue bulk} feature. 
Dotted lines indicate other possible solutions described in the text.}
\end{figure}

\section{Discussion and conclusions}
The best agreement between the simulations and the data is obtained by adopting the SFH 
plotted in Figure~\ref{sfh} (solid line). 
Keeping in mind that these results are affected by many uncertainties (both 
in  data and models), 
we can derive the following overall scheme for the SFH of NGC 1569
(Recall that the results described in this paper refer to 
the central region of the galaxy).
\begin{itemize}
 \item [-] NGC 1569 has experienced a complex SFH, 
composed of at least three strong episodes in the last 1 (possibly 2) Gyr.
 \item [-] The quiescent phase that occurred $(1.5 - 3) \times 10^8$ yr ago is inferred 
from the small number of  $3 - 6$ \Msun \ stars in the AGB phase.
 \item [-] We need an episode older than $3 \times 10^8$ yr to reproduce the fainter
part of the CMD. This old SF may start as early as 1 $-$ 2 Gyr ago.
 The large uncertainties (severe incompleteness, large photometric errors) that 
characterize the fainter part of the CMD do not allow us to infer reliable details on the SF 
of the oldest episode. However, we find hints for a significantly lower SF activity 
(or even a quiescent phase) at ages older than 1 $-$ 2 Gyr 
(for a distance of 2.2 Mpc).
 \item [-] A Salpeter IMF reproduces the brighter part of the observed CMD 
and the LF reasonably well. A flatter IMF is also able to reproduce the 
brighter portion of the red plume.
 \item [-] By assuming a Salpeter IMF over the whole mass range, the most recent 
SF occurs at a rate of $3.1 - 3.3$ \Myk \ ($0.12 - 0.14$ \Myr), the
intermediate one at a rate of $1 - 1.2$ \Myk \ ($0.03 -0.05$ \Myr), and
the older one at a rate between  0.25 and 1.5 \Myk \ ($0.01-0.06$ \Myr),
depending on the assumed duration.
 \item [-] The uncertainties of the distance modulus  affect the estimate
of the ages and SFRs of the episodes. 
The best fit obtained with a larger distance of $\mathrm{DM} = 27.3$ is 
$2.7 \times 10^7 - 8 \times 10^6$ yr ago for the young and 
$1 \times 10^8 - 3.2 \times 10^7$ yr ago for the intermediate episode.
The corresponding SFRs are about 2 times higher than those found for $\mathrm{DM} = 26.71$.
A SF with a rate of $0.8$ \Myk \ (0.04 \Myr) occurred in the age interval 
$1.3 \times 10^9 - 2 \times 10^8$ yr ago.
Also for the larger distance, we can exclude a strong SF that occurred at ages older 
than 1 $-$ 2 Gyr because the RGB-Tip would be clearly present in the CMD.
 \item [-] We cannot exclude more than one SF episodes occurring
in the age interval $(1 - 4) \times 10^7$ yr . A possible scenario could consist of three episodes 
with a SFR of 4.2, 4.5 and 3.2 \Myk \ (0.18, 0.19 and 0.13 \Myr) 
separated by short phases of quiescence or low activity. The uncertainties are due to  
the possible presence of clusters at the top of the red plume and small number statistics.
\end{itemize}

% Match with SFH from other simulation
By comparing our results and those from the literature for the SFH of NGC 1569,
we find overall agreement, although with interesting differences.
The end of the youngest episode ($1.3 \times 10^7$ yr ago) is significantly older than that obtained 
by \citet{val96} and G98, who place it around $(5 - 10) \times 10^6$ yr ago, as found from
optical HST data. These differences can be explained in terms of a different 
sensitivity of the different bands (optical vs. NIR) to hot and cold evolutionary phases. 
Here we are studying young massive stars with NIR data. We are then sensitive to the 
post-main-sequence phases, where these stars have very short lifetimes. Optical
data, more appropriate for main-sequence objects, allow for higher reliability
on the youngest stars. 
On the other hand, the NIR analysis allows us to distinguish the two recent 
episodes, separated by a short quiescent period, better than with the optical data.
In addition, we find evidence for a gap, or a low SF from
$3 \times 10^8$ yr ago to $1.5 \times 10^8$ yr ago, similar to that suggested by \citet{val96}.

\begin{figure}
 \figurenum{9}
 \includegraphics[bb=18 520 385 718, width=0.48\textwidth]{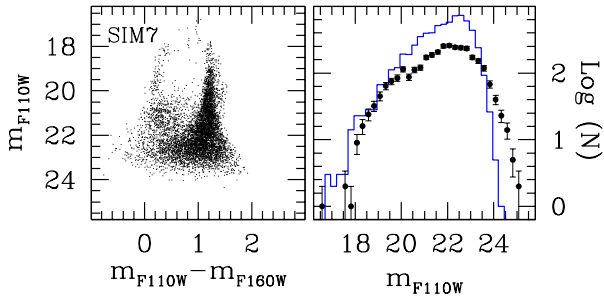}
 \label{Gre}
 \figcaption{Simulation of an episode with duration and SFR as found by G98. 
These parameters do not allow us to reproduce the observed CMD and LF.}
\end{figure}

 The SFR derived by G98 using the PC camera (which covers an area of $374 \times 374$ pc$^2$ for 
a distance of 2.2 Mpc), is $4$ \Myk \ in the age interval $1.5\times 10^8 - 5\times 10^6$ yr ago. 
From Fig.~\ref{sfh}, we can see that our rate for the youngest episode (SIM4)
is slightly lower than that derived by G98. For the intermediate episode,
we have derived a SFR $3 - 4$ times lower than G98.
 We performed further simulations by adopting the 
ingredients suggested by the literature cited above in order to better understand these differences. 
We cannot reproduce the top of the red plume if we simulate the totality of objects in the CMD 
with a continuous and constant activity 
starting $(1 - 1.5) \times 10^8$ yr ago and ending $5 \times 10^6$ yr ago.  Furthermore, this 
simulation underestimates the fainter part of the LF.  
Another simulation (SIM7, plotted in Fig.~\ref{Gre}) assumes the SFR of G98 in the same age interval.
The resulting synthetic diagram has twice as many stars than
 our observed CMD. The total luminosity function is broadly fitted in the brighter part of the 
red plume but shows an overabundance of objects in the range \mj \ $= 19.5 - 23.5$, 
whereas the fainter part is barely populated. 
From the comparison of the observed LF with that from SIM7, we can learn that:
\begin{itemize}
 \item[-] If we terminate the SF $5\times 10^6$ yr ago, we produce many high-mass 
stars without observed counterparts;
 \item[-] In the BOX1 area, SIM7 produces a number of objects $\sim 1.5$ times higher than SIM4.
Nevertheless, the SFRs of SIM7 and SIM4 are quite similar within the uncertainties since the length of the two 
simulations is slightly  different.
 \item[-] The simulated objects of SIM7 in the BOX2 region are about $3-4$ times higher than SIM5 
(this value is similar to the ratio of the corresponding SFRs). This discrepancy could arise from 
the uncertainties in the photometric error distribution and completeness factors used in this 
work and in G98. In addition, the differences in the SFH could result from the differences in the 
areas sampled by the two cameras.
\end{itemize}

% Understanding SFH of NGC 1569 
%% SF in Cluster
By using archival HST data, \cite{and04} find a high-level of 
star cluster formation for ages younger than $2.5 \times 10^7$ yr and a lower 
level of star cluster formation at older epochs, with a secondary peak about 100 Myr ago. 
The cluster age distribution shows a gap in the age interval $(1.6 - 4.0) \times 10^8$ yr. 
These results match our SFH that refers to field stellar population.
However, \citet{and04} find a younger age for the end of the last episode,
consistent with the results by  \citet{val96} and G98.  This again suggests that the optical 
data are more appropriate for studying the most recent epochs.

% Heckman 95
In order to estimate whether the expanding superbubbles can be plausibly 
driven by the observed population of massive stars, \citet{hec95} found 
that the observations can be reproduced reasonably well by assuming an episode with a 
constant SF over the ages $1.3 \times 10^7 -  3.2 \times 10^7$ yr ago.
The ages of the episode are rather similar to our youngest burst.

% SF from X-ray binary
 By using population synthesis models, \cite{bel04} have constructed synthetic
X-ray binary populations for direct comparison with the X-ray LF and found two 
principal stellar populations.
One is old, metal-poor, with continuous SF from 1.5 Gyr ago and the other is recent
and metal-rich. Their results are consistent with ours for the oldest episode, 
but invoke a recent burst younger than that inferred here. 
Again, we attribute our earlier termination to the selection effect which 
makes it difficult to detect blue young stars in the NIR. We remind the reader that
the most recent termination epoch for the SF inferred in the optical by G98 only refers
to the regions of NGC 1569 where single stars were resolved. There are
zones in this galaxy occupied by giant HII region complexes where even with
the HST/WFPC2 we were unable to resolve individual stars. The existence  of these
HII regions clearly suggests that the SF activity must have continued until even more recent times.

All three episodes have SFRs higher by 2--3 orders of
magnitude than that of the solar neighborhood [e.g. \citet{tim95}] 
and than those inferred with similar methods in other late-type
dwarfs [both irregulars and BCDs, see the reviews by \citet{sch01}
 and by \citet{tos03}]. The only dwarf analyzed so far with the synthetic CMD
method which has a SFR comparable to that of NGC 1569 is NGC 1705
\citep{ann03}. These two dwarfs, in spite of being  classified differently
(NGC 1705 is a BCD), have quite similar properties, all somehow related to the
strong SF activity. They both contain SSCs,  show evidence of 
galactic winds, and have SFRs consistent with those required by
\citet{bab96} to let dwarfs  account for the excess of blue galaxies at
intermediate redshift. Neither shows evidence of long quiescent phases
in the last couple of Gyrs.

 At the rates derived in this paper, the central region of NGC 1569 has formed
$\sim 2.7 \times 10^7$ \Msun \ of stars in the last 1 Gyr,
i.e. about $8\%$ of the total mass estimated by \cite{isr88} and about $14\%$ of the mass that
is not associated with neutral hydrogen ($M_{dif} = 2 \times 10^8$ \Msun).
The field of view of NIC2 corresponds to an area of 0.04 kpc$^2$, which covers only 
$\sim 2.5\%$ of the optical size of NGC 1569 [1.76 kpc$^2$ \citep{isr88}]. 

To improve our knowledge of the SF history we have compared the integrated
optical and infrared fluxes of the whole field and of the resolved stars
in the NICMOS area, from the images described by G98 an snd AL01.
The integrated magnitudes in the HST F439W, F555W, F110W and F160W
filters are 13.7, 12.9, 10.9 and 10, respectively, for the
field;\footnote{These values do not include the flux from the SSCs and from the
HII regions, which were masked on the frames.} they are 15.2, 14.5, 12.1, and 11 
for the resolved stars. The dereddened m$_{\mathrm{F439W}}-$ m$_{\mathrm{F160W}}$ color
of our investigated region is then $\simeq$ 1.6, noticeably
bluer than the corresponding color for a solar Kurucz model
\citep[][(m$_{\mathrm{F439W}} -$ m$_{\mathrm{F160W}}$)$_{\odot} \simeq$ 2.2]{kur92}.
 The apparent integrated B magnitude of our NICMOS field is
m$_{\mathrm{B,NICMOS}} \simeq 13.6$, whilst the total B magnitude of NGC~1569 is
m$_{\mathrm{B}} = 11.86$  \citep{dev91}. Hence, our NICMOS frame samples
$\sim$ 20\% of the total B light from the galaxy, if the average reddening
of the NICMOS region is representative of that of the whole galaxy.
Notice that our sampled area is only 2.5\% of the total, therefore
the SF per unit area in NGC~1569 is very centrally concentrated.

We can explore some possible scenarios in order to place constraints on the
past history of NGC 1569.
Given the total and HI masses of NGC 1569 estimated by \cite{isr88} and the
astrated mass in the last 1 Gyr extracted from our CMD procedure and assuming a
Salpeter's IMF over the whole mass range, we are left
with only  $1.73 \times 10^8$ \Msun \ for the whole galaxy for the sum of the older stars and the dark
matter. Assuming that these old stars were produced at a constant rate over the
interval  $1.3 \times 10^{10} - 1 \times 10^9$ yr ago, 
the corresponding SFR is $1.44 \times 10^{-2}$ \Myr. Since this SFR refers to the whole galaxy, 
the normalized value is $8.2 \times 10^{-3}$ \Myk. This value is comparable to what has been derived 
via simulations for dwarf galaxies in the Local Group (see G98 and references therein).
For the NIC area, the corresponding astrated mass over the age interval is 
$M_{NIC} = 4.3 \times 10^6$ \Msun.
If we simulate an episode that generates the $M_{NIC}$ mass over the age 
interval reported above, we obtain a synthetic CMD that contains $5 - 10\%$ of the 
total objects of the observed CMD. These synthetic stars are on the RGB.
Hence, in this scenario we cannot rule out that a constant but low SF 
might have occurred over the Hubble time. 

However, it would be very unusual for a galaxy of this type to have a dark/visible 
mass ratio much lower than 10. Moreover, chemical evolution models \citep{rom05}
show that without the usual dark matter ratio, NGC 1569 would experience too
early strong galactic winds and it would be impossible to reconcile the model
predictions with the observational data.

In the second scenario (Fig. 8  dashed line) we therefore assume
that the stellar component is only 10\% of the dynamical mass; then our astrated
masses are probably inconsistent with a Salpeter's IMF extrapolated down to 0.1
\Msun. However if we adopt \citet{kro03} IMF, the central region has generated only $\sim 60\%$ 
of the stellar mass of the galaxy over the last 1 Gyr.
The resulting mean SFR over the interval $1.3 \times 10^{10} - 1 \times 10^9$ yr ago is 
$6.7 \times 10^{-4}$ \Myk. Alternatively, considering an onset of the SF 7 Gyr ago 
(i.e., the galaxy started to form stars at redshift 1 by adopting the ``concordance 
cosmology'' [$H_0=70$, $\Omega_{\lambda}=0.7$, $\Omega_{\mathrm{M}}=0.3$]), the mean SFR is about 
2 times higher than the value above. 
 All these quantities lie in the range of SFRs typical for nearby dwarf galaxies.
These scenarios do not allow us to rule out a past SF with 
a low rate (or similar to the typical SFR of the other dwarf galaxies), but we can safely
conclude that the last 1 $-$ 2 Gyr have been peculiar. 

What could be the cause of this peculiar SF in the last Gyr?
Interestingly, there is an HI cloud with mass  $7 \times 10^6$ M$_{\odot}$ 
located at a projected distance of 5 kpc from NGC 1569. It is connected to the galaxy 
by a bridge similar in mass to the HI cloud \citep{sti98}. 
This cloud may have triggered the star formation \citep{sti98}. 
Moreover, according to Muehle (2003), who studied the
galaxy with high-resolution HI data, the structure of the galaxy
halo is the remnant of an intergalactic HI cloud in the process of
merging with the galaxy. If true, this may provide both the cause and
the fuel for the strong and concentrated SF activity of NGC 1569.

To better understand the history of this intriguing object, we should try to
resolve its stars older than the few Gyr reached by our current data, in both
the inner and outer regions. 
Although the NIR bands are important to explore old stellar population, 
NICMOS is not an optimal instrument.
The detection of the tip of the RGB (if it exists) will be achievable only 
with the HST Advanced Camera for Surveys, which has both the necessary 
spatial coverage and resolution.

%%%%%%%%%%%%%%%%%%%%%%%%%%%%%%%%%%%%%%%%%%%%%%%%%%%%%%%%%%%%%%%%%%%%%%%%%%%%%%%
%%                                                                           %%
%%                                                                           %%
%%                       A C K N O W L E D G M E N T S                       %%
%%                                                                           %%
%%                                                                           %%
%%%%%%%%%%%%%%%%%%%%%%%%%%%%%%%%%%%%%%%%%%%%%%%%%%%%%%%%%%%%%%%%%%%%%%%%%%%%%%%
\acknowledgments{}
We thank Francesca Annibali, Cristian Vignali, Paolo Montegriffo, 
Livia Origlia, Marcella Maio and Lucia Pozzetti for their help and useful 
discussions.
The authors also thank the anonymous referee for useful suggestions.
This work has made use of the NASA Astrophysics Data System, 
the arXiv.org e-print archive, and the NASA/IPAC Extragalactic Database.
This work has been partially supported by the Italian ASI, through grant 
2002-IT059, and MIUR, through Cofin 2000.

%%%%%%%%%%%%%%%%%%%%%%%%%%%%%%%%%%%%%%%%%%%%%%%%%%%%%%%%%%%%%%%%%%%%%%%%%%%%%%%
%%                                                                           %%
%%                                                                           %%
%%                        B I B L I O G R A P H Y                            %%
%%                                                                           %%
%%                                                                           %%
%%%%%%%%%%%%%%%%%%%%%%%%%%%%%%%%%%%%%%%%%%%%%%%%%%%%%%%%%%%%%%%%%%%%%%%%%%%%%%%

\clearpage

\end{document}